 \documentclass[preprint2,longabstract]{aastex}

\shorttitle{The Cepheid in the binary system OGLE-LMC-CEP-2532}
\shortauthors{Pilecki, Graczyk, Gieren et al.}

\begin{document}


\title{The Araucaria Project: The First-Overtone Classical Cepheid in the Eclipsing System OGLE-LMC-{\allowbreak}CEP-2532
\footnote{The research is based on observations obtained with the ESO VLT and 3.6 m telescopes for Programmes 092.D-0295(A), 091.D-0393(A), 089.D-0330(A), 088.D-0447(A), 086.D-0103(A) and 085.D-0398(A)), and with the Magellan Clay and Warsaw telescopes at Las Campanas Observatory.}
}


\author{Bogumi{\l} Pilecki}
\affil{Warsaw University Observatory, Al. Ujazdowskie 4, 00-478, Warsaw, Poland}
\affil{Universidad de Concepci{\'o}n, Departamento de Astronomia, Casilla 160-C, Concepci{\'o}n, Chile}
\email{pilecki@astrouw.edu.pl}
\author{Dariusz Graczyk}
\affil{Millenium Institute of Astrophysics, Santiago, Chile}
\affil{Universidad de Concepci{\'o}n, Departamento de Astronomia, Casilla 160-C, Concepci{\'o}n, Chile}
\email{darek@astro-udec.cl}
\author{Wolfgang Gieren}
\affil{Universidad de Concepci{\'o}n, Departamento de Astronomia, Casilla 160-C, Concepci{\'o}n, Chile}
\affil{Millenium Institute of Astrophysics, Santiago, Chile}
\email{wgieren@astro-udec.cl}
\author{Grzegorz Pietrzy{\'n}ski}
\affil{Warsaw University Observatory, Al. Ujazdowskie 4, 00-478, Warsaw, Poland}
\affil{Universidad de Concepci{\'o}n, Departamento de Astronomia, Casilla 160-C, Concepci{\'o}n, Chile}
\email{pietrzyn@astrouw.edu.pl}
\author{Ian B. Thompson}
\affil{Carnegie Observatories, 813 Santa Barbara Street, Pasadena, CA 91101-1292, USA}
\email{ian@obs.carnegiescience.edu}
\author{Rados{\l}aw Smolec}
\affil{Copernicus Astronomical Centre, Polish Academy of Sciences, Bartycka 18, 00-716 Warsaw, Poland}
\email{smolec@camk.edu.pl}
\author{Andrzej Udalski}
\affil{Warsaw University Observatory, Al. Ujazdowskie 4, 00-478, Warsaw, Poland}
\email{udalski@astrouw.edu.pl}
\author{Igor Soszy{\'n}ski}
\affil{Warsaw University Observatory, Al. Ujazdowskie 4, 00-478, Warsaw, Poland}
\email{soszynsk@astrouw.edu.pl}
\author{Piotr Konorski}
\affil{Warsaw University Observatory, Al. Ujazdowskie 4, 00-478, Warsaw, Poland}
\email{piokon@astrouw.edu.pl}
\author{M{\'o}nica Taormina}
\affil{Universidad de Concepci{\'o}n, Departamento de Astronomia, Casilla 160-C, Concepci{\'o}n, Chile}
\email{mtaormina@astro-udec.cl}
\author{Alexandre Gallenne}
\affil{Universidad de Concepci{\'o}n, Departamento de Astronomia, Casilla 160-C, Concepci{\'o}n, Chile}
\email{agallenne@astro-udec.cl}
\author{Dante Minniti}
\affil{Departamento de Ciencias F{\'i}sicas, Universidad Andr{\'e}s Bello, Rep{\'u}blica 220, Santiago, Chile}
\affil{Millenium Institute of Astrophysics, Santiago, Chile}
\email{dante@astrofisica.cl}
\author{Marcio Catelan}
\affil{Instituto de Astrof{\'i}sica, Facultad de F{\'i}sica, Pontificia Universidad Cat{\'o}lica de Chile, Av. Vicu{\~n}a MacKenna 4860, Santiago, Chile}
\affil{Millenium Institute of Astrophysics, Santiago, Chile}
\email{mcatelan@astro.puc.cl}

\begin{abstract}
We present here the first spectroscopic and photometric analysis of the double-lined eclipsing binary containing the classical, first-overtone Cepheid OGLE-LMC-CEP-2532 (MACHO 81.8997.87). The system has an orbital period of 800 days and the Cepheid is pulsating with a period of 2.035 days.

Using spectroscopic data from three high-class telescopes and photometry from three surveys spanning 7500 days we are able to derive the dynamical masses for both stars with an accuracy better than $3\%$. This makes the Cepheid in this system one of a few classical Cepheids with an accurate dynamical mass determination ($M_{1}=3.90 \pm 0.10 M_{\odot}$). The companion is probably slightly less massive ($3.82 \pm 0.10 M_{\odot}$), but may have the same mass within errors ($M_{2}/M_{1}= 0.981 \pm 0.015$). The system has an age of about 185 million years and the Cepheid is in a more advanced evolutionary stage.

For the first time precise parameters are derived for both stars in this system. Due to the lack of the secondary eclipse for many years not much was known about the Cepheid's companion. In our analysis we used extra information from the pulsations and the orbital solution from the radial velocity curve. The best model predicts a grazing secondary eclipse shallower than 1 mmag, hence undetectable in the data, about 370 days after the primary eclipse.

The dynamical mass obtained here is the most accurate known for a first-overtone Cepheid and will contribute to the solution of the Cepheid mass discrepancy problem. 
\end{abstract}


\section{Introduction}  
\label{sect:intro}

Detached binary systems and classical Cepheids are important astrophysical tools to measure distances and to study the evolution and the physical properties of stars. Since the discovery of the Cepheid period-luminosity relation (Leavitt 1908) Cepheid variables have been important distance indicators in the local universe (e.g. Gieren et al. 2005, Freedman \& Madore 2010, Riess et al. 2011, Kodric et al. 2015). They are also key objects for testing the predictions of stellar evolution and stellar pulsation theory (Caputo et al. 2005). Although significant progress has been made in the understanding of the physics of Cepheid stars, the lack of an independent and accurate measurement of the dynamical masses has meant that the “Cepheid mass discrepancy problem” has persisted for more than 40 years (Cox 1980, Keller 2008, Evans et al. 2008, Neilson et al. 2011 and references therein).

Detached double-lined eclipsing binary systems are excellent tools to measure the masses of the stars (Andersen 1991, Torres et al. 2010). Masses can be measured with an accuracy better than $1\%$ even at distances of the Large \citep{pie13} and Small \citep{gra14} Magellanic Clouds. Yet for many years these two classes of objects were studied separately because no Cepheid in an eclipsing double-lined system was known.
This situation has started to change with the microlensing surveys that monitored millions of stars. Several candidates of Cepheids in eclipsing binaries were announced (Alcock et al. 2002, Soszyński et al. 2008) however none of them could be confirmed as a double-lined physical system from high resolution spectroscopy.   The breakthrough came with the pioneering work of \cite{pie10} on the classical Cepheid OGLE-LMC-CEP-0227 and since then a number of other Cepheids in eclipsing binaries have been spectroscopically confirmed \citep{pie11,gra13,gie14}.
As the methodology has improved and more data are gathered it becomes possible to measure the radii and masses of Cepheids with an accuracy of $1\%$ as shown by \cite{pil13}, a reanalysis of the OGLE-LMC-CEP-0227 system.

Interest in Galactic binary Cepheids has been also recently revived thanks to the spatial detection of high-contrast companions from long-baseline interferometry (LBI) and adaptive optics (AO) imaging (Gallenne et al. 2014a, 2014b and references therein). So far, only one Galactic
Cepheid -- Polaris -- has its mass measured (Evans et al. 2008), but with the use of AO and LBI, it will be possible to combine astrometric motions with radial velocities to obtain an independent estimate of the mass and distance of Galactic Cepheids.

The binary nature of OGLE-LMC-CEP-2532 (LMC-SC16-119952 in the OGLE project database) was first noticed by Udalski et al. (1999). One eclipse was detected superimposed on the classical overtone Cepheid variability of 2.035 days.  The star was later identified as object 81.8997.87 in the MACHO project database. Alcock et al. (2002) analyzed combined OGLE and MACHO light curves and derived an orbital period of 800.5 days. Using a simple geometrical model with a circular orbit they concluded that the companion to the Cepheid is a much fainter red giant branch star of spectral type K or later. The system was subsequently reanalyzed by Lepischak et al. (2004) using a similar model but allowing for eccentric orbits. Despite of the lack of radial velocity data and the absence of a secondary eclipse they were able to obtain some basic information (like the relative radii, intrinsic colors of the stars and the inclination) with a reasonable accuracy. They also studied different possibilities regarding the nature of the components, but the conclusion was consistent with the findings of the previous authors.
Finally, OGLE-LMC-CEP-2532 was again identified as a first-overtone (FO) Cepheid showing eclipses by Soszyński et al. (2008) in their catalog of classical Cepheids in the Large Magellanic Cloud (LMC).

The OGLE-LMC-CEP-2532 system was then included in the Araucaria Project\footnote{http://araucaria.astrouw.edu.pl} program of precise measurements of the parameters of Cepheids in eclipsing binary systems. First results of our follow-up were given by \cite{gra13} who reported that the system was a genuine double-lined eclipsing binary on a significantly eccentric orbit. The orbit and the masses of the components were still poorly constrained but the resulting mass ratio was consistent with equal masses for the Cepheid and the secondary star. The binary lies within the 30 Dor star-forming region and the Cepheid component is substantially fainter than a typical overtone Cepheid in the LMC with a period of 2 days. This result was interpreted as evidence for the presence of strong interstellar extinction in this direction (Alcock et al. 2002).

In this work we present a full analysis of the OGLE-LMC-CEP-2532 system using the same methodology as used by Pilecki et al. (2013) for the OGLE-LMC-CEP-0227 system. We use spectroscopic data from three 4-8m class telescopes and photometry from three surveys spanning 7500 days. We present the orbital solution and the absolute physical parameters of both components. The evolutionary status of the stars is also briefly discussed.

\section{Data}
\label{sect:data}

\begin{deluxetable}{cccccccccccc}
\tabletypesize{\scriptsize}
\tablecaption{Radial velocity measurements of both components of the OGLE-LMC-CEP2532 system. HARPS, UVES, and MIKE blue and red data are marked with H, U and M$_b$, M$_r$, respectively.}
\tablewidth{0pt}
\tablehead{
\colhead{HJD} & \colhead{RV$_1$} & \colhead{RV$_2$} & \colhead{Instr.} & \colhead{HJD} & \colhead{RV$_1$} & \colhead{RV$_2$} & \colhead{Instr.} & \colhead{HJD} & \colhead{RV$_1$} & \colhead{RV$_2$} & \colhead{Instr.}
}
\startdata
5430.8224 & 254.49 & 297.19 & H & 5883.8272 & 303.77 & 254.71 & M$_{r}$ & 6179.8103 & 248.64 & 299.23 & U \\
5457.8952 & 254.99 & 291.71 & M$_{b}$ & 5884.7661 & 289.08 & 254.35 & U & 6179.8513 & 246.89 & 299.67 & U \\
5457.8953 & 256.41 & - & M$_{r}$ & 5884.7996 & 288.26 & 254.98 & U & 6181.8870 & 246.18 & 300.46 & U \\
5458.8750 & 272.03 & - & M$_{b}$ & 5884.8287 & 287.65 & 255.22 & U & 6572.7705 & 280.88 & - & H \\
5458.8750 & 273.81 & 292.73 & M$_{r}$ & 5885.6853 & 299.90 & 255.16 & U & 6579.8817 & 298.11 & 259.80 & M$_{b}$ \\
5477.7171 & 253.04 & 289.18 & H & 5885.7318 & 300.64 & 255.38 & U & 6579.8817 & 298.40 & 260.42 & M$_{r}$ \\
5478.7932 & 270.41 & 290.19 & H & 5885.7654 & 301.55 & 254.93 & U & 6637.7441 & 295.83 & 256.30 & M$_{r}$ \\
5502.6929 & 261.07 & 286.81 & H & 5886.6947 & 295.76 & 254.38 & U & 6637.7441 & 296.80 & 256.50 & M$_{b}$ \\
5503.6406 & 275.63 & - & H & 5886.7403 & 290.60 & 254.32 & U & 6638.6020 & 295.51 & 254.61 & M$_{b}$ \\
5503.8098 & 274.29 & 287.06 & H & 5886.7738 & 289.17 & 254.28 & U & 6638.6020 & 295.74 & 254.86 & M$_{r}$ \\
5833.7923 & 292.75 & 257.57 & M$_{r}$ & 5886.8045 & 287.91 & 254.45 & U & 6666.6749 & 288.41 & 256.59 & M$_{b}$ \\
5833.7923 & 292.57 & 254.56 & M$_{b}$ & 5886.8351 & 286.99 & 254.53 & U & 6666.6749 & 288.63 & 254.94 & M$_{r}$ \\
5836.7919 & 298.93 & 256.61 & M$_{b}$ & 5951.5596 & 300.60 & 261.32 & M$_{b}$ & 6669.6003 & 304.72 & 254.80 & M$_{r}$ \\
5836.7920 & 297.55 & 255.72 & M$_{r}$ & 5951.5603 & 300.55 & 261.20 & M$_{r}$ & 6674.8391 & 287.76 & 256.01 & M$_{r}$ \\
5861.8127 & 307.33 & 255.29 & U & 5952.7103 & 291.50 & 263.13 & M$_{r}$ & 6931.7915 & 250.51 & 303.52 & M$_{b}$ \\
5862.8022 & 289.91 & 255.73 & U & 5952.7104 & 291.01 & 262.62 & M$_{b}$ & 6931.7917 & 250.99 & 304.17 & M$_{r}$ \\
5862.8385 & 290.33 & 255.00 & U & 5965.5949 & 298.20 & 262.53 & M$_{b}$ & 6932.8471 & 250.46 & 302.16 & M$_{b}$ \\
5871.8337 & 306.22 & 255.33 & U & 5965.5949 & 299.23 & 265.11 & M$_{r}$ & 6932.8471 & 252.00 & 303.13 & M$_{r}$ \\
5873.6743 & 301.60 & 252.92 & U & 6162.8699 & 259.06 & 301.01 & U & 7018.7189 & 244.68 & 297.29 & M$_{b}$ \\
5873.7772 & 304.23 & 254.36 & U & 6174.8186 & 256.30 & 299.85 & U & 7018.7189 & 245.11 & 298.04 & M$_{r}$ \\
5873.8149 & 305.01 & 254.82 & U & 6174.8836 & 257.23 & 299.73 & U &  \\
5883.8272 & 303.09 & 253.62 & M$_{b}$ & 6175.8589 & 243.90 & 300.79 & U &  
\label{tab:rvdata}
\enddata
\end{deluxetable}

As the object is very faint and lacks the secondary eclipse it was very important to obtain a high-quality radial velocity curve. Moreover the pulsations of the Cepheid and its period of 2.035 days makes the analysis even harder, as the orbital motion has to be separated from the radial velocity changes due to the pulsations. In general one needs at least 30 well spaced measurements. Using data from three high-class instruments we were able to disentangle the radial velocities and obtain both a high-quality orbital, and pulsational radial velocity curve of the Cepheid. 

We obtained 49 high-resolution spectra using the UVES spectrograph (at 25 epochs) at the ESO Very Large Telescope located at Paranal, the MIKE spectrograph (17 epochs) at the 6.5-m Magellan Clay telescope at Las Campanas and the HARPS spectrograph (9 epochs) attached to the ESO 3.6-m telescope at La Silla.

Velocities were measured using the Broadening Function method (Rucinski 1992, 1999) implemented in the RaveSpan software (Pilecki et al. 2012). Measurements were made in the wavelength interval 4125 to 6800 $\mathring{A}$. Spectra taken from the
library of Coelho et al. (2005) were used as templates. The typical formal errors of the derived velocities are $\sim800$ m/s. The measurements are listed in Table~\ref{tab:rvdata} and shown later in the Section~\ref{sect:results}. In some cases where line profiles of both companions were blended, only the velocity of the Cepheid (number 1 in the table) was measured. We assumed that all three spectrographs are on the same
velocity system.

A total of 1322 photometric measurements in the $I$-band and 231 in the $V$-band were collected with the Warsaw telescope by the OGLE project (Udalski 2003) and during time granted to the Araucaria project by the Chilean National Time Allocation Comittee (CNTAC). We have also used instrumental $V$-band and $R$-band data from the MACHO project downloaded from the webpage http://macho.anu.edu.au and converted to the Johnson-Cousins system using equations from Faccioli et al.(2007). We augmented our data with $R_{EROS}$ (equivalent to Johnson-Cousins I) data from the EROS project (Kim et al. 2014)\footnote{http://stardb.yonsei.ac.kr}. The EROS identification of this system is lm0034l19415. During a literature search we also found data for this system obtained by Lepischak et al. (2004), but because of the small number of out-of-eclipse observations we could not merge these univocally with the rest of our data and they were not included in the analysis.

\begin{table}[hbt!]
\caption{Coordinates and observed out-of-eclipse mean magnitudes of CEP-2532} 
\begin{tabular}{llc}
\hline
Coordinates & Right Asc. & Declination\\
Epoch 2000.0 & 05h36m04.46s &$-70^{\rm o}$01'55.7" \\
 \hline
Band     &   Mag   &  Reference      \\
\hline
$<B>$    &  18.67  &  Udalski+(2000) \\ 
$<V>$    &  17.29  &  This paper     \\
$<R_C>$  &  16.30  &  This paper     \\ 
$<I_C>$  &  15.74  &  This paper     \\ 
$<K_s>$  &  13.72  &  Ripepi+(2012) \\ 
\hline
 \label{tab:phot}
\end{tabular}
\end{table}

\begin{figure}[hbt!]
\begin{center}
  \resizebox{\linewidth}{!}{\includegraphics{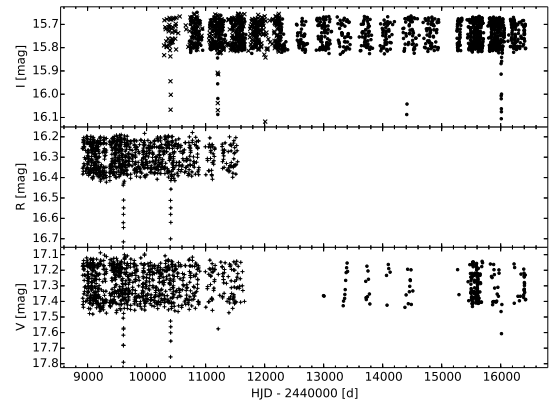}} \\
\caption{Photometric data collected for OGLE-LMC-CEP-2532. {\it Upper panel:} OGLE (dots) and converted EROS-R (x) $I$-band data; {\it middle panel:} MACHO $R$-band data; {\it lower panel:} OGLE (dots) and MACHO (crosses) $V$-band data.}
\label{fig:obs_all}
\end{center}
\end{figure}

The basic photometric information is given in Table~\ref{tab:phot} and all of the photometry used in this work is shown in Fig.~\ref{fig:obs_all}. The photometic and radial velocity observations are available online  at:

\centerline{ http://araucaria.astrouw.edu.pl/p/cep2532 }

\begin{figure}[hbt!]
\begin{center}
  \resizebox{\linewidth}{!}{\includegraphics{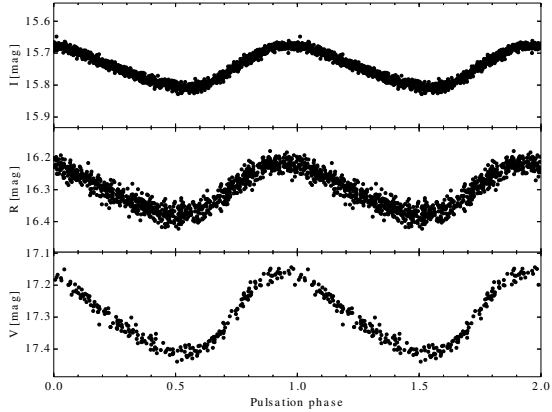}} \\
\caption{Out-of-eclipse light curves of the OGLE-LMC-CEP-2532 folded with the ephemeris $T_{max} ($HJD$) = 2454507.1019 + 2.03534862\times $E. The pulsational light curves contain the light of the companion. Y-axis span is always the same (0.4 mag) -- one can see that the amplitude is smaller for the redder filters.
\label{fig:lcpuls}}
\end{center}
\end{figure}

Out-of-eclipse light curves of OGLE-LMC-CEP-2532 folded with the pulsation period are plotted in Fig.~\ref{fig:lcpuls}. A quick analysis of the shape of the $I$-band light curve of the Cepheid component through its Fourier decomposition parameters confirmed that the star is a first-overtone pulsator as already stated by Udalski et al. (1999) and Lepischak et al. (2004). We stress that at pulsation periods around 2 days the Fourier decomposition parameters firmly discriminate first-overtone and fundamental mode pulsators (Soszynski et al. 2008).

\section{Results}
\label{sect:results}

From an analysis of the radial velocity curve of a binary star one can obtain the orbital parameters of the system. The situation is complicated in this case by the pulsational variability of the Cepheid superimposed on the orbital motion. Using the RaveSpan software we have fitted a model of Keplerian orbit (i.e. proximity effects were not included) with an additional Fourier series representing the pulsational radial velocity curve of the Cepheid.

We simultaneously fitted the reference time $T_0$, the eccentricity $e$, the argument of periastron $\omega$, the velocity semi-amplitudes $K_1$ and $K_2$, the systemic velocities $\gamma_1$ and $\gamma_2$, and 8th-order Fourier series. The period $P$ was initially held fixed at the estimated value of 800.4 days. The fitting was later repeated with fixed values of $P=800.4191$ and $T_0=5199.697$ as obtained from the analysis of the photometry to ensure the consistency of the model.

\begin{table}[hbt!]
\caption{Orbital solution for OGLE-LMC-CEP-2532; $T_0$ ($HJD - 2450000$~d) calculated from the epoch of the primary minimum: 5210.8233.} 
\begin{tabular}{lr@{ $\pm$ }lc}
 \hline
 Parameter & \multicolumn{2}{c}{Value} & Unit \\
 \hline 
$\gamma_1$      &  277.84   &  0.50  &  km/s   \\ 
$\gamma_2$      &  275.55   &  0.32  &  km/s   \\
$T_0$           & \multicolumn{2}{c}{5199.697 (fixed)} &  d      \\
$a \sin i$      &  715.6    &  5.9   &  $R_\odot$ \\ 
$m_1 \sin^3 i$  &    3.87   &  0.10  &  $M_\odot$ \\ 
$m_2 \sin^3 i$  &    3.80   &  0.10  &  $M_\odot$ \\ 
$q=m_2/m_1$     &    0.981  &  0.014 &  $M_\odot$  \\ 
$e$             &    0.307  &  0.012 &  -      \\
$\omega$        &  100.1    &  2.7   &  deg  \\
$K_1$           &   23.53   &  0.26  &  km/s \\ 
$K_2$           &   23.99   &  0.22  &  km/s \\ 
rms$_1$         & \multicolumn{2}{c}{0.8} &  km/s \\
rms$_2$         & \multicolumn{2}{c}{0.87} &  km/s\\
\hline
 \label{tab:spec}
\end{tabular}
\end{table}

\begin{figure}[hbt!]
\begin{center}
  \resizebox{\linewidth}{!}{\includegraphics{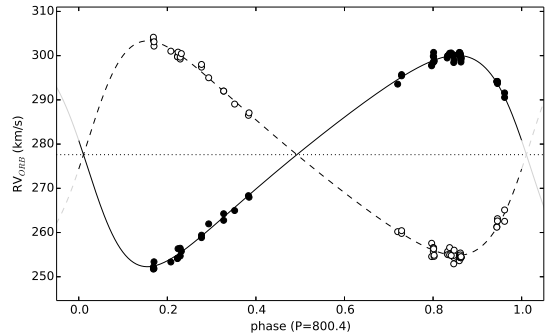}} \\
\caption{Orbital solution for OGLE-LMC-CEP-2532. Measured radial velocities of the Cepheid with the pulsations removed (filled circles) and of its non-pulsating companion (open circles) are presented.
\label{fig:rvorb}}
\end{center}
\end{figure}

\begin{figure}[hbt!]
\begin{center}
  \resizebox{\linewidth}{!}{\includegraphics{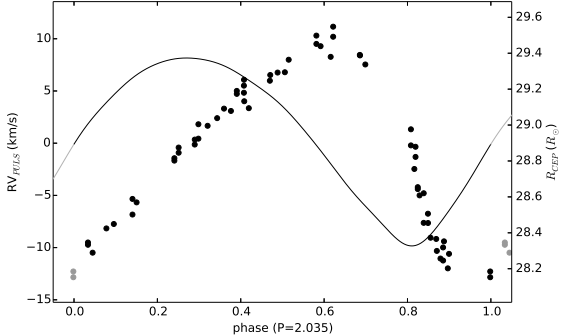}} \\
\caption{Pulsational radial velocity curve (points) and the radius variation of the Cepheid over one pulsation cycle (solid line). To obtain this RV curve the orbital motion was removed from the measured radial velocities. The full amplitude of the radius change is 1.05 R$_\odot$. Repeated data points outside the 0.0 -- 1.0 phase range are shown in grey. 
\label{fig:rvpuls}}
\end{center}
\end{figure}

In this way we have obtained the coefficients describing the pulsational radial velocity curve and the parameters describing the orbital motion separately. The orbital radial velocity curve along with the best fitted model is shown in Fig.~\ref{fig:rvorb}.
To obtain the radial velocity curve of the Cepheid we then subtracted the orbital motion from the measured velocities. This pulsational curve is shown in Fig.~\ref{fig:rvpuls} together with the radius change calculated with an assumed $p$-factor 1.2. The orbital solution is presented in Table~\ref{tab:spec}.


The photometric data were analyzed using a version of the JKTEBOP code (Popper \& Etzel 1981, Southworth 2004, 2007) modified to allow the inclusion of pulsation variability. We have previously used this package in the analysis of the OGLE-LMC-CEP-0227 system \citep[hereafter P13]{pil13}, and we refer the reader to this work for more details.
However, the OGLE-LMC-CEP-2532 system is fainter (and thus the quality of the data is lower), and there is no evidence for a secondary eclipse in the data. This makes the system more complicated to analyze and forced us to introduce some simplifications.

{\it First}, during the analysis we kept the eccentricity and the argument of periastron fixed as the lack of a secondary eclipse makes it impossible to get a reliable determination of these parameters from the light curve alone. We adopted values from the orbital solution for these parameters.

{\it Second}, the value of the $p$-factor was held fixed. The small radius change of the first-overtone Cepheid (see Fig.~\ref{fig:rvpuls}) means that the effect on the light curve is not  significant given the quality of the data. We adopted a value of 1.2 (as derived for the  Cepheid OGLE-LMC-CEP-0227, see P13), but within the expected range its effect on the results is negligible. In fact, when we tried to include the parameter in the analysis without any constraints we obtained a value $p=0.7 \pm 1.4$. Any realistic value (1.1-1.5) is well within 1-$\sigma$ range. The full amplitude of the radius change of the Cepheid is $1.05  R_\odot$ for an assumed $p=1.2$.

{\it Third}, we did not take third light into account. There is no indication of any in the collected spectra, and  the quality of the photometry and the lack of the secondary eclipse preclude a reliable analysis assuming the presence of third light.

{\it Fourth}, the limb darkening  coefficients for the secondary component were calculated for a temperature of 5000 K, but as the star is not eclipsed, this has no effect on the results. For the Cepheid we tried limb darkening  coefficients for a range of effective temperatures between 3500 and 7000 K taking into account the low value obtained for OGLE-LMC-CEP-0227 (P13). However the results did not change significantly with the maximum difference in the $\chi^2$ between the best models being 0.3. For the final model we adopted a temperature of 6500 K, a value  consistent with that calculated in Section~\ref{subs:abs}. For all calculations we used stellar limb darkening tables published by Van Hamme (1993) and a logarithmic law as indicated by the results of analysis in P13.

In general, with only a primary eclipse visible it is hard to reliably determine the radii of the components and the inclination of the system. In this case, however, there are some factors which helped us to obtain a well defined model. The orbital parameters suggest that the secondary eclipse should appear at phase 0.46. Our photometry data rule out any eclipse deeper than 0.003 mag in the $I$-band at this phase. This puts a tight upper limit for the orbital inclination and the sum of the radii -- a higher inclination or a larger sum of the radii would make the secondary eclipse deeper and visible. Moreover, the pulsational variability influences the eclipse shape by the change of the radius and the temperature of the eclipsed component.  Our analysis does not remove the pulsations from the light curves and so we obtain additional information in comparison with a singe-eclipse binary system of two stable stars.
And finally, we made a simultaneous modeling of the light curves in three different passbands assuming the same system geometry. This improved the quality of the results by removing some correlations between model parameters, there is a significantly smaller number of parameter sets that simultaneously match the three different light curves.

\begin{deluxetable}{lccl}
\tablecaption{Photometric parameters of OGLE-LMC-CEP-2532 from the Monte Carlo simulations. Asterisk-marked values correspond to a pulsation phase 0.0, epoch of the primary eclipse $T_{I}$ is $HJD - 2450000$~d. $L_{21}$ is the light ratio of the components in every photometric band.}
\tablewidth{0pt}
\tablehead{
\colhead{Parameter} & \colhead{Mean value} & \colhead{Fitted value} & \colhead{Stat. error}
}
\startdata
$P_{orb} (d) $  & -       & 800.419  & 0.009  \\
$T_{I}$ (d)     & -       & 5210.823  & 0.033\\
$r_1$           & 0.0404  & 0.0403$^*$ & 0.0019 \\
$r_2$           & -       & 0.0526   & 0.0024 \\
$j_{21}(V)$     & 0.292   & 0.244$^*$ & -0.039 / +0.023 \\
$j_{21}(R_C)$   & 0.449   & 0.393$^*$ & -0.046 / +0.027 \\ 
$j_{21}(I_C)$   & 0.465   & 0.415$^*$ & -0.047 / +0.020 \\
$i$ ($^\circ$)  & -       & 85.97     & -0.24 / +0.08 \\
$e$             & -       & 0.307     & fixed \\
$\omega$ ($^\circ$)& -    & 100.1     & fixed\\
$p$-factor      & -       & 1.20      & fixed\\
\multicolumn{4}{l}{Derived quantities:}\\
$r_1+r_2$       & 0.0931  & 0.0930$^*$ & 0.0011 \\
$L_{21}(V)$     & 0.49    & 0.42$^*$ & -0.09 / +0.07 \\ 
$L_{21}(R_C)$   & 0.76    & 0.67$^*$ & -0.12 / +0.10\\ 
$L_{21}(I_C)$   & 0.79    & 0.71$^*$ & -0.13 / +0.10\\ 
\multicolumn{4}{l}{Additional information:}\\
rms ($V$)       & 0.030   &  &\\ 
rms ($R_C$)     & 0.018   &  &\\
rms ($I_C$)     & 0.009   &  &
\label{tab:photpar}
\enddata
\end{deluxetable}


\begin{figure*}
\begin{center}
  \resizebox{\linewidth}{!}{\includegraphics{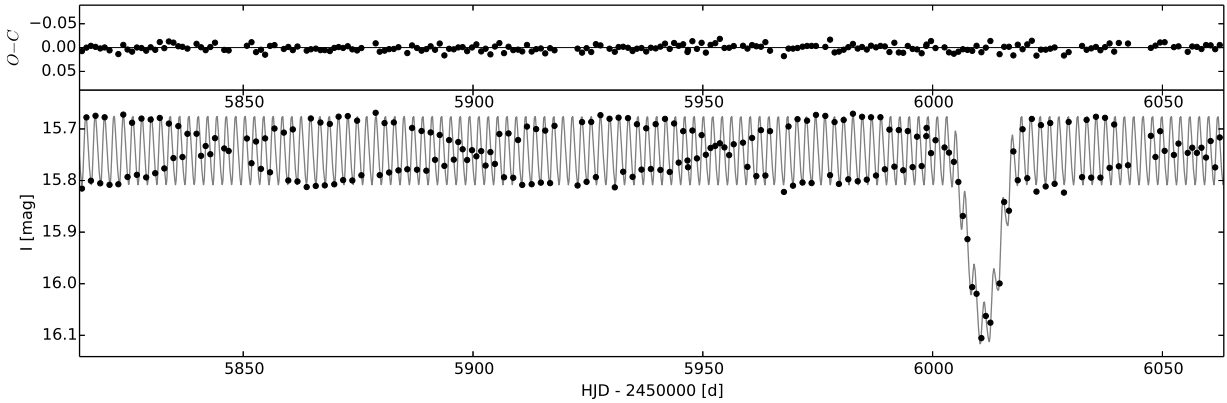}} \\
\caption{$I$-band OGLE-IV data (points) and the model (solid line). No systematics in residuals is visible in the eclipse. The apparent beating of the observations is caused by the pulsating period being almost equal to 2 days.
\label{fig:ilongmodel}}
\end{center}
\end{figure*}


\begin{figure*}
\begin{center}
  \resizebox{0.32\linewidth}{!}{\includegraphics{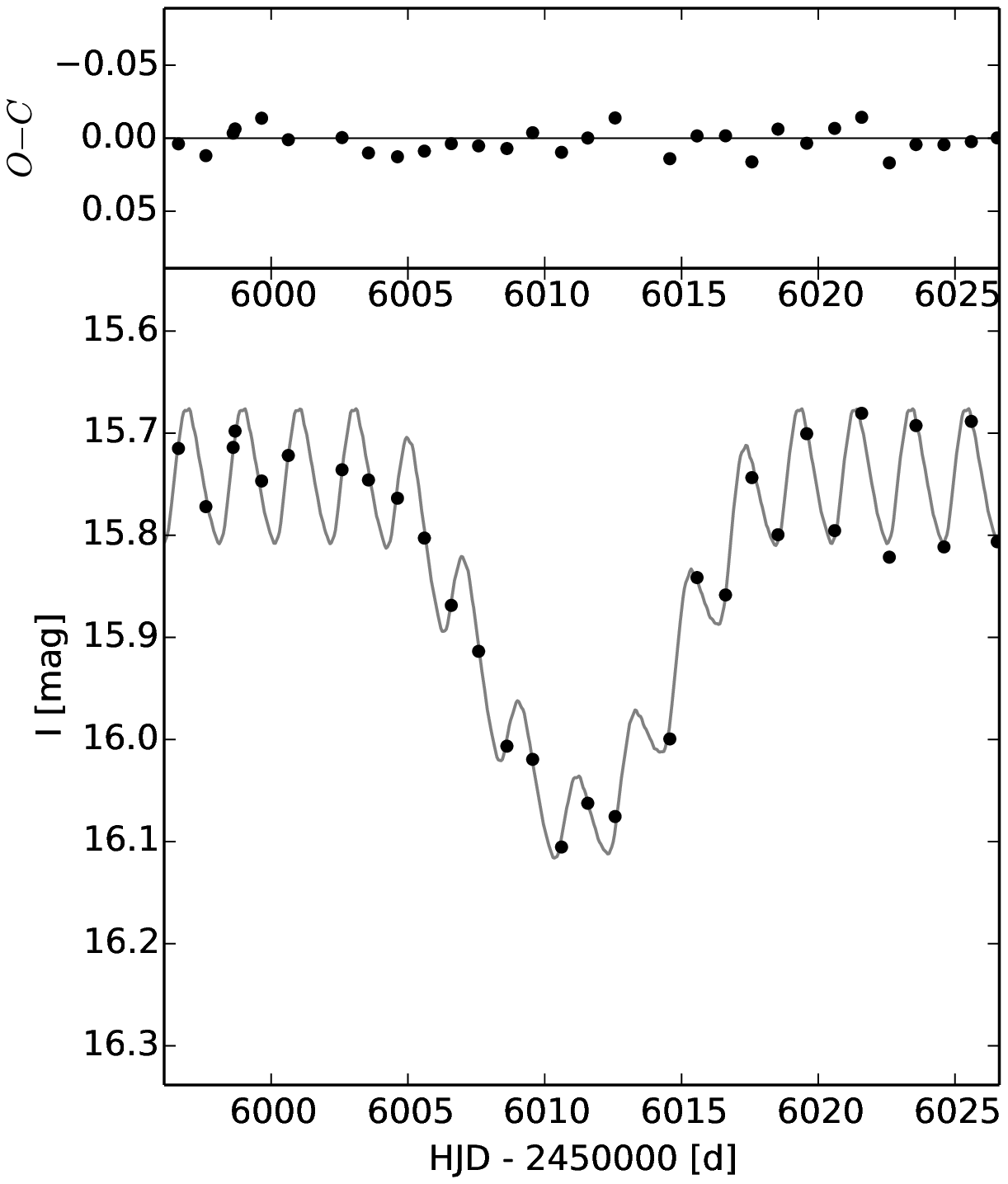}}
  \resizebox{0.32\linewidth}{!}{\includegraphics{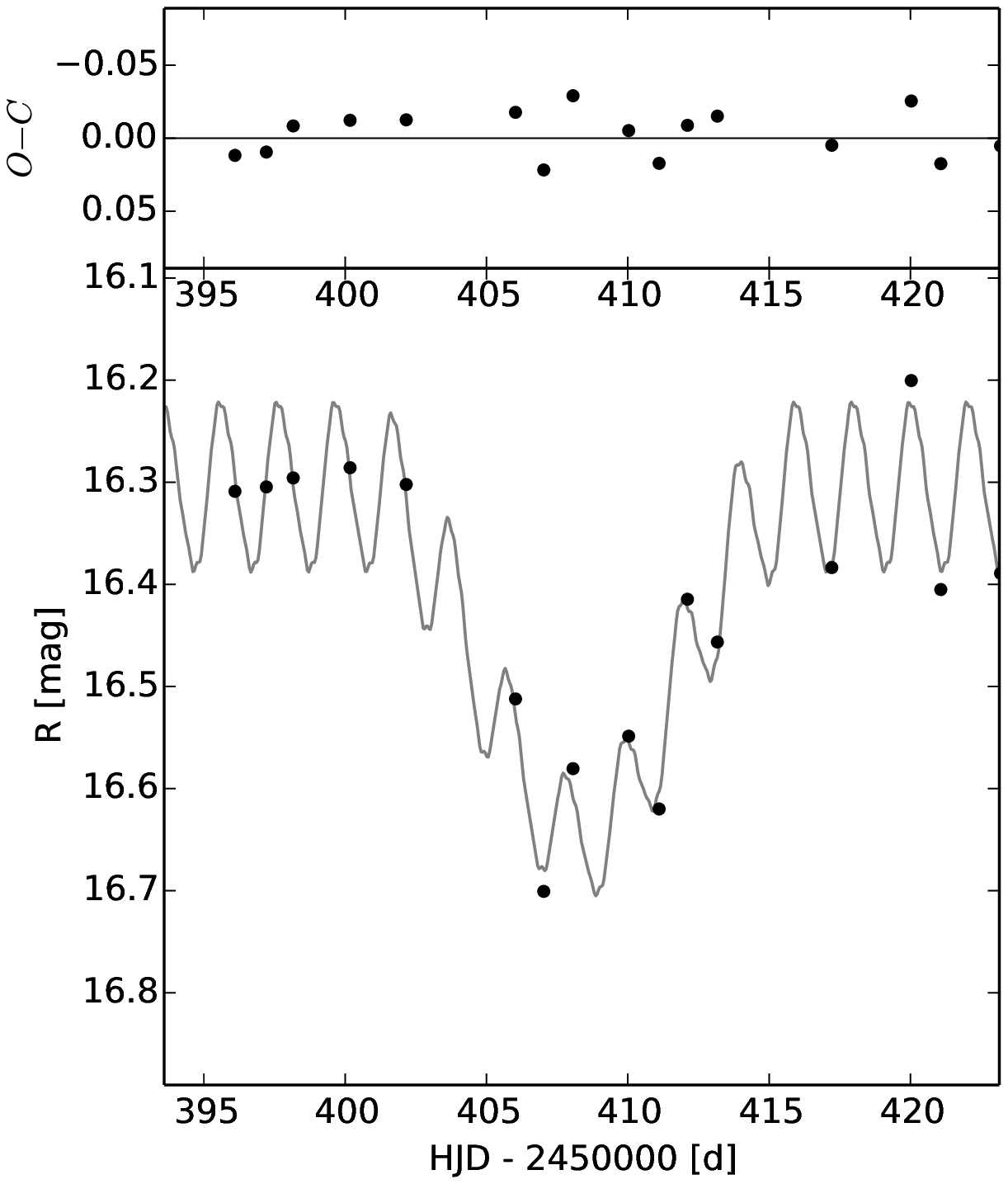}}
  \resizebox{0.32\linewidth}{!}{\includegraphics{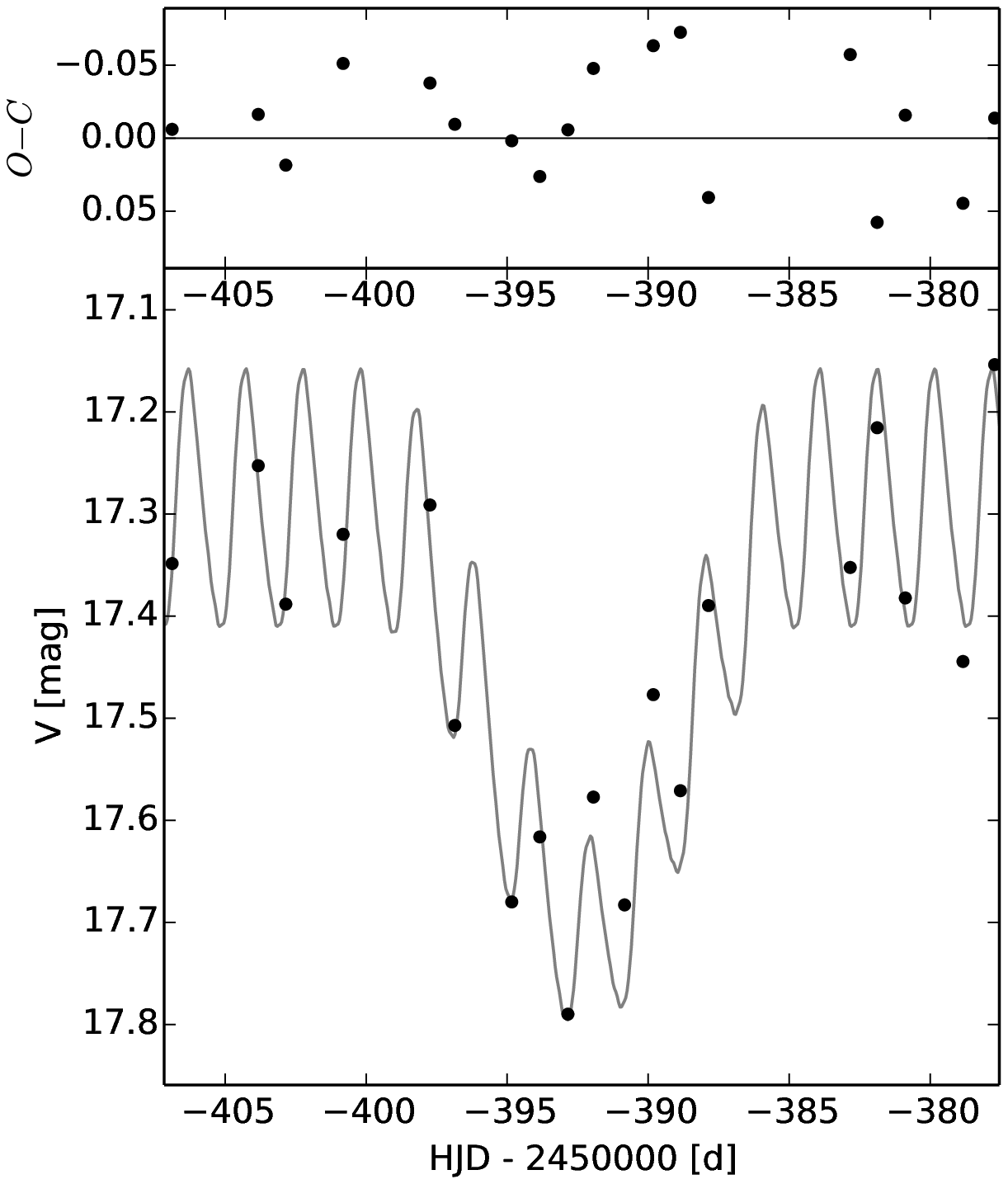}}
\caption{Close-up on the selected eclipses. The eclipse is caused by the transit of the companion star over the disc of the Cepheid. As expected the pulsation amplitude is lower during eclipses due to the smaller contribution of the Cepheid to the total light. The {\em rms} is about 0.01, 0.02 and 0.03 for $I_C$, $R_C$ and $V$, respectively.
\label{fig:ieclmodel}}
\end{center}
\end{figure*}

We varied the following parameters in deriving the final model: the fractional radius of the pulsating component at phase 0.0 (pulsational) -- $r_1$, the fractional radius of the second component -- $r_2$, the orbital inclination $i$, the orbital period -- $P_{orb}$, the epoch of the primary minimum -- $T_{I}$ and the component surface brightness ratios in all three photometric bands at phase 0.0 (pulsational) -- $j_{21}$. The radius change of the Cepheid was calculated from the pulsational radial velocity curve using the assumed $p$-factor value and the change of the surface brightness ratios from the instantaneous radii and out-of-eclipse pulsational light curves (for details, see P13).
The search for the best model (lowest $\chi^2$ value) was made using the Markov chain Monte Carlo (MCMC) approach (Press et al. 2007) as described in P13. The best fit photometric parameters are presented in Table~\ref{tab:photpar}. Using these parameters we generated a model for each light curve. A portion of the $I$-band data is presented in Fig.~\ref{fig:ilongmodel}, no systematics are visible in any part of the light curve. In Fig.~\ref{fig:ieclmodel} we show a close-up of selected eclipses for each passband. The amplitude is larger toward blue wavelengths, and the quality of the data is the best for the $I$-band data and the worst for the $V$-band data.

\begin{figure}[hbt!]
\begin{center}
  \resizebox{\linewidth}{!}{\includegraphics{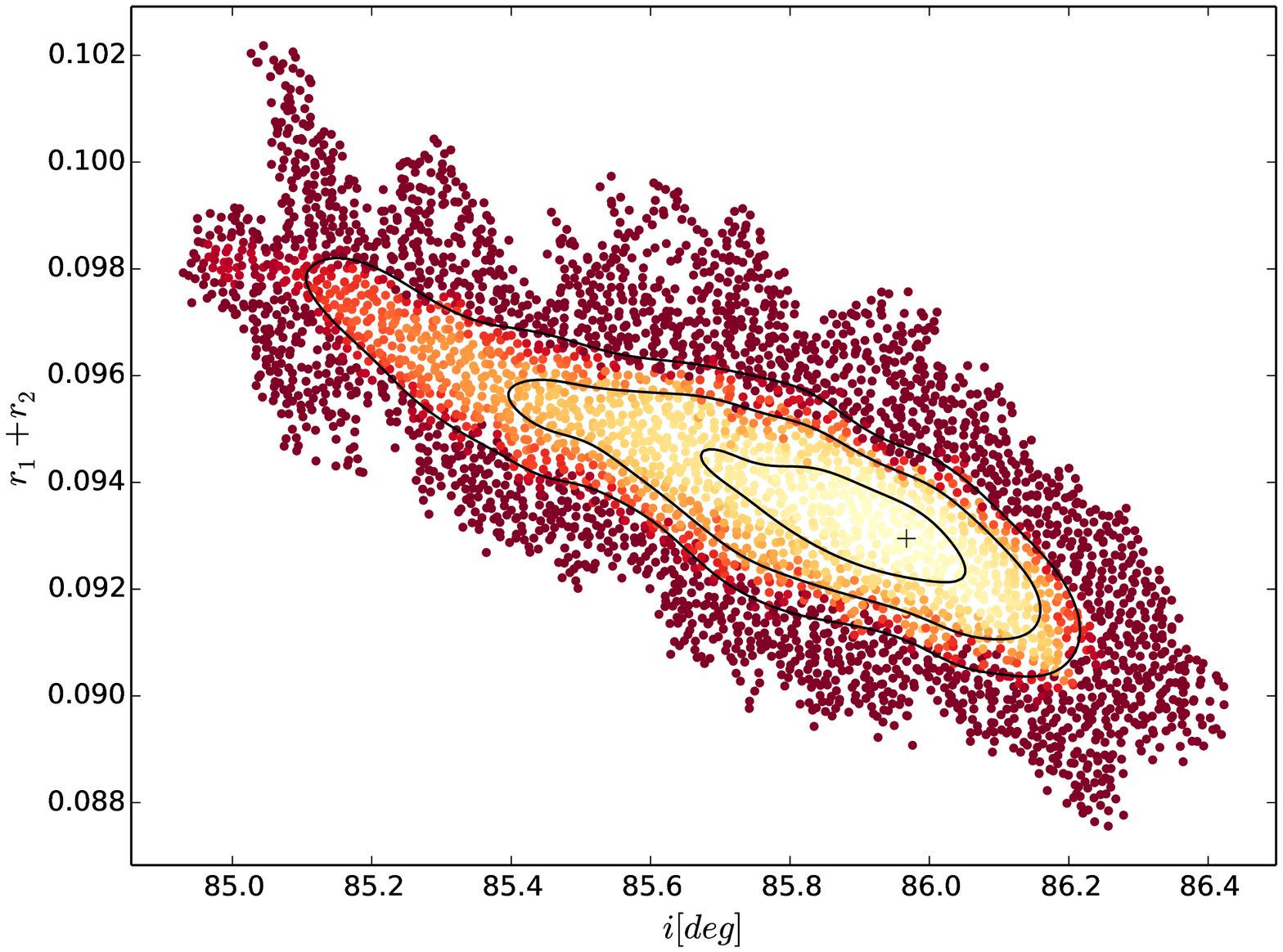}} \\
  \resizebox{\linewidth}{!}{\includegraphics{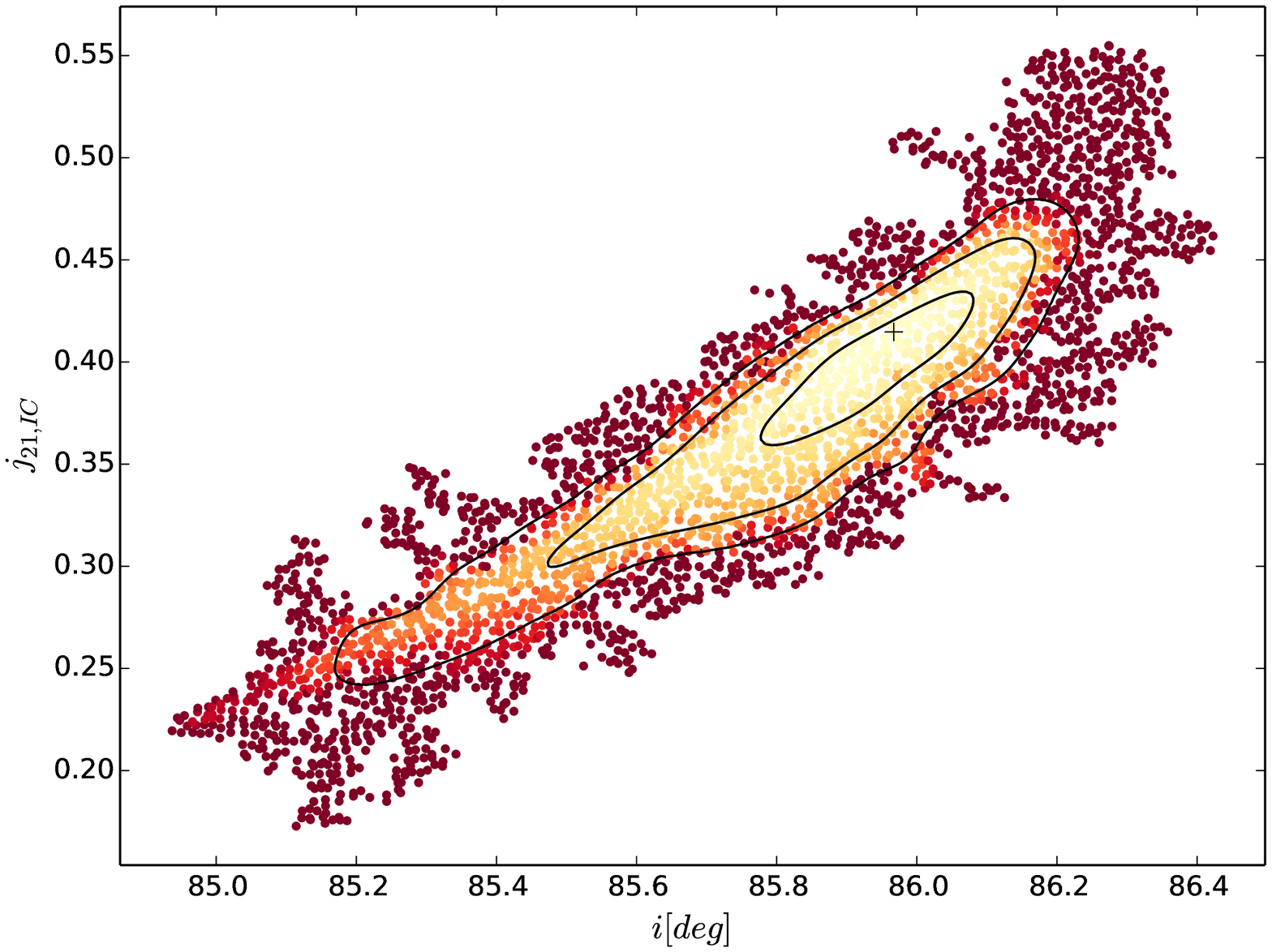}} \\
\caption{Correlation between the inclination and the sum of the star fractional radii (top) and between the inclination and the surface brightness ratio in $I$-band (it is similar for other bands). The $\chi^2$ values are coded with color (higher values are darker). Solid lines represent 1-, 2- and 3-$\sigma$ levels for the two-parameter error estimation. Best model is marked with a cross.
\label{fig:corr_i}}
\end{center}
\end{figure}

Most of the parameters fitted in our approach are independent and do not exhibit any significant correlation. With the eccentricity and the $p$-factor fixed the only significant correlation is between the inclination and the surface brightness ratios, between the orbital plane inclination $i$ and the sum of the radii $r_{1}+r_{2}$ and between the radii themselves (almost a perfect anti-correlation). The first two are shown in Fig.~\ref{fig:corr_i}. Note that because of the correlation the sum of the radii is much better constrained that the individual radii themselves (see Table~\ref{tab:photpar}).

\subsection{Eclipses}
\begin{figure}[hbt!]
\begin{center}
  \resizebox{\linewidth}{!}{\includegraphics{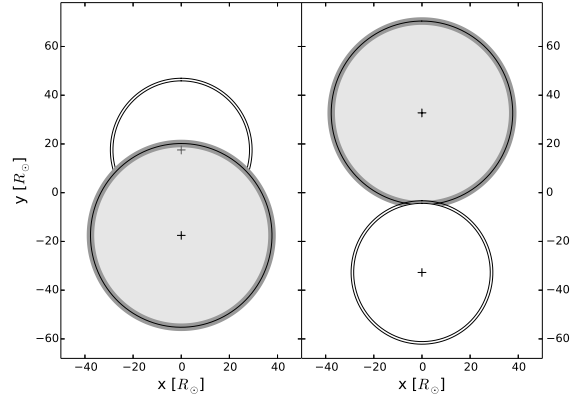}} \\
\caption{System configuration at the primary ({\em left}) and the secondary eclipse ({\em right}). For the~Cepheid (white) the minimum and maximum radius is shown. For the companion (light grey) the 1-$\sigma$ radius error is marked with dark-grey color (it is of the same order for the Cepheid but is not shown for clarity). The distances between the stars are about 500 $R_\odot$ and 930 $R_\odot$ at the primary and secondary eclipse, respectively. Star centers are marked with crosses.
\label{fig:config}}
\end{center}
\end{figure}

To better understand the configuration of the system using the derived parameters we calculated the distances between the stars at important phases. At the phase of the primary eclipse the distance between the components is about $500 R_{\odot}$, while at the phase of the secondary eclipse it is about $930 R_{\odot}$. This difference, together with the orbital inclination, is responsible for the apparent lack of a secondary eclipse. However, calculating the projected distance between the centers of the stars shows that a grazing eclipse is possible. At the primary eclipse this distance is $35.1 R_{\odot}$, and at the secondary the distance  is $65.5 R_{\odot}$, while the sum of the radii changes between 66.1 and 67.2 $R_{\odot}$ depending on the radius of the Cepheid. The configuration at both phases is illustrated in Fig.~\ref{fig:config}.

\subsection{Reddening and Absolute Dimensions}
\label{subs:abs}
Within the JKTEBOP code there is no need to specify the temperatures of the components of an eclipsing binary. However, when using tabular limb darkening coefficients in light curve modeling some estimate of temperature is needed. Initially we assumed an effective temperature of the primary component of 5000 K. After obtaining the first photometric solution we derived temperatures of both components as follows. Using surface brightness ratios $j_{21}(V)$ and $j_{21}(I_C)$ and relative radii of components $r_1$, $r_2$ we calculated light ratios in two bands:  $L_{21}(V)$ and $L_{21}(I_C)$ (see Table~\ref{tab:photpar}).  As a first step in deriving the reddening of the system we calculated the differential (specific minus LMC average) extinction as follows. Utilizing the observed (reddened) period-luminosity relations for FO Cepheids in the LMC (Soszyński et al. 2008) we derived the expected observed mean magnitudes of the Cepheid in the $V$-band and $I_C$-band for a pulsation period of P=2.035 $d$. We then calculated the magnitudes of the companion star using previously derived light ratios and combined the light of both stars to get the expected mean magnitude of the system in the $V$-band and $I_C$-band. Comparison of these magnitudes with the observed mean magnitudes of the system (see Table~\ref{tab:phot}) showed a high degree of reddening, consistent with the results of Alcock et al. (2002). Subtracting expected magnitudes from the observed values we obtained the differential extinction in both bands -- see Table~\ref{tab:extin}. 

The derived values for the extinction were converted into a color excess E(B-V) assuming $R_V=3.1$ and the ratio of extinction in both bands $A_I/A_V$=0.59.
We obtained similar E(B-V) from each band and we calculated an average of E(B-V)=0.53$\pm 0.06$ mag, where the error takes into account the uncertainty of the light ratio of the components and the internal spread of period-luminosity relations for FO Cepheids. This reddening does not include foreground and mean internal reddening of the LMC. The mean extinction of the LMC is $A_V=0.38$ (Imara\&Blitz 2007), corresponding to E(B-V)=0.122 mag. The total color excess in the direction of OGLE-LMC-CEP-2532 is thus E(B-V)=0.65$\pm 0.08$ mag. This reddening is significantly larger then the reddening derived by Alcock et al. (2002), but we benefit from a more precise magnitude disentangling of the components in two bands and the longer wavelength baseline between V and I. The reddening corresponds to a total extinction of $A_V=2.03$ mag and $A_I=1.2$ mag. Combining the dereddened magnitudes of the system and the light ratios we then derived an individual magnitude for each component. Absolute magnitudes were derived assuming the distance modulus from Pietrzynski et al.(2013) -- see Table~\ref{tab:extin}.

Ripepi et al. (2012) reported a preliminary dereddened period luminosity relation for classical Cepheids in the LMC for the $K$-band. From their relation we derived a dereddened magnitude of the Cepheid component of $K=14.68$ mag. Using our measured reddening  and the relation $A_K=0.34\cdot E(B-V)$ we estimated a total extinction in the $K$-band as $A_K=0.21$ mag. The mean observed magnitude of the OGLE-LMC-CEP-2532 system reported by Ripepi et al. (2012) is $K=13.72$ mag, and thus the dereddened magnitude of the companion is $K=13.94$ mag  (see Table~\ref{tab:extin}). Finally, from the individual magnitudes we derived intrinsic (V-K) and (V-I) colors (see Table~\ref{tab:abs}). These colors were used to calculate temperatures and bolometric corrections of the components using the calibration of Worthey \& Lee (2011). We obtain consistent temperatures (within 120 K) using both colors suggesting that the estimate of the total extinction is reliable. Average values from both colors are 6467 K and 4660 K for the Cepheid and the companion, respectively. The (V-K) and (V-I) colors of the companion are consistent with spectral type K1 $\pm$ 1 subtype according to the calibration by Bessell \& Brett (1988).  

\begin{deluxetable}{lcccccccc}
\tabletypesize{\scriptsize}
\tablecaption{Magnitudes of OGLE-LMC-CEP-2532. Flux-weighted means are given for the Cepheid and the system (total). Differential extinction is marked with $\Delta$.}
\tablewidth{0pt}
\tablehead{
 \colhead{} & \colhead{magnitude} & \colhead{expected} & \colhead{observed} & \multicolumn{2}{c}{extinction} & \colhead{dereddened} & \colhead{absolute$^{a}$} & \colhead{ bolometric$^{e}$}\\
 \colhead{} & \colhead{from P-L} & \colhead{magnitude} & \colhead{magnitude} & \colhead{$\Delta$} & \colhead{total} & \colhead{magnitude} & \colhead{magnitude} & \colhead{magnitude}
}
\startdata 
  \multicolumn{9}{c}{$V$-band}  \\
total      		&  -  		& 15.64	& 17.29	&1.65	&2.03$^{f}$&15.26	&-		& -  \\ 
Cepheid        &16.06$^{b}$& 16.06 	& 17.72	&-		&-		&15.68	&-2.81	&-2.81 \\
companion      	&   - 		&16.86$^{d}$& 18.51&-		&-		&16.48	&-2.01	&-2.36  \\ 
\hline
\multicolumn{9}{c}{$I_C$-band}  \\
total      		&  -  		& 14.76	& 15.74	&0.98	&1.20$^{f}$&14.54	&-		& - \\ 
Cepheid        &15.39$^{b}$& 15.39 	& 16.37	&-		&-		&15.17	&-3.32	& - \\
companion      	&   - 		&15.66$^{d}$& 16.64&-		&-		&15.44	&-3.05	& - \\ 
\hline
\multicolumn{9}{c}{$K$-band}  \\
total      		&  -  		& -	& 13.72	&-		&0.22$^{g}$&13.50	&-		& - \\ 
Cepheid        &14.68$^{c}$& - 	& 14.90	&-		&-		&14.68	&-3.81	& - \\
companion      	&   - 		& -	& 14.17	&-		&-		&13.94	&-4.55	& -
\label{tab:extin}
\enddata
\tablenotetext{a}{ from distance modulus to the LMC of 18.49 (Pietrzynski et al. 2013)}
\tablenotetext{b}{ observed (reddened) relation for FO classical Cepheids from Soszy{\'n}ski et al. (2008)}
\tablenotetext{c}{ extinction corrected (dereddened) relation for FO classical Cepheids from Ripepi et al. (2012)}
\tablenotetext{d}{ from light ratio of the components}
\tablenotetext{e}{ bolometric corrections from Worthey\&Lee (2011)}
\tablenotetext{f}{ added foreground and mean internal reddening of the LMC: $A_V=0.38$ mag and $A_I=0.22$ mag}
\tablenotetext{g}{ from relation $A_K=0.34\cdot E(B-V)$}
\end{deluxetable}

The absolute dimensions of the components are summarized in Table~\ref{tab:abs}. Using individual bolometric magnitudes (Table~\ref{tab:extin}) and the absolute bolometric magnitude of the sun $M_{\odot}=+4.83$ mag we derived the luminosities of the components. 
Combining this information with the absolute radii we obtained effective temperatures of 6223 K and 4945 K
for the components. These temperatures are slightly different from those obtained from color indices. Finally we adopted an average of both temperature estimates. 

\begin{deluxetable}{lcc}
\tablecaption{Physical properties of OGLE-LMC-CEP-2532. The spectral type, (mean) radius, gravity ($\log g$), temperature, luminosity ($\log L$) and the dereddened magnitudes and colors are mean values over the pulsation cycle. The orbital period is a rest frame value.}
\tablewidth{0pt}
\tablehead{
 \colhead{Parameter} & \colhead{Primary (Cepheid)} & \colhead{Secondary}}
\startdata
spectral type &  F4 II & K1 III-II \\
mass ($M_\odot$) & 3.90 $\pm$ 0.10 & 3.83 $\pm$ 0.10 \\ 
radius ($R_\odot$) & 28.95 $\pm$ 1.4  & 37.7 $\pm$ 1.7 \\ 
$\log g$ (cgs) & 2.11 $\pm$ 0.04 & 1.87 $\pm$ 0.04  \\ 
temperature (K) & 6345$\pm$ 150 & 4800 $\pm$ 220 \\ 
$\log L$ ($L_\odot$) & 1219 $\pm$ 164 & 677 $\pm$ 138 \\
$V$ (mag) & 15.68 & 16.48\\
(V-$I_C$) (mag) & 0.51 & 1.04\\
(V-K) (mag) & 1.00 & 2.54\\
E(B-V) (mag) &\multicolumn{2}{c}{$0.65 \pm 0.08$}\\
orbital period (days) & \multicolumn{2}{c}{799.678 $\pm$ 0.009 } \\
semimajor axis ($R_\odot$) & \multicolumn{2}{c}{716.8 $\pm$ 5.9}
\label{tab:abs}
\enddata
\end{deluxetable}

Table~\ref{tab:abs} presents the physical parameters of both components. The error in the estimate of the Cepheid mean radius is of the same order as the total radius change, this makes an independent determination of the $p$-factor for the OGLE-LMC-CEP-2532 Cepheid impossible with the current data. The spectral type of the Cepheid was estimated from the effective temperature using the calibration given by \cite{alo99}.

\subsection{Evolutionary status}

\begin{figure}[hbt!]
\centering
\resizebox{\hsize}{!}{\includegraphics{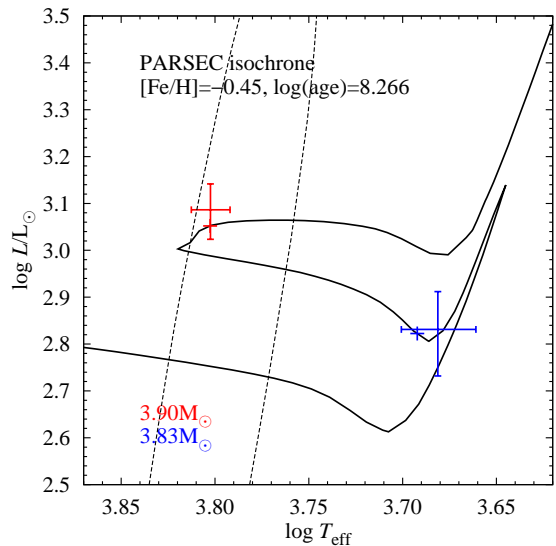}}
\caption{PARSEC isochrone for ${\rm [Fe/H]}=-0.45$ at log age=8.266. Small crosses along the isochrone mark the points with masses corresponding to the OGLE-LMC-CEP-2532 masses. Dashed lines mark the linear edges of tthe instability strip for first overtone Cepheids as computed in Baranowski et al. (2009).}
\label{fig:isochrone}
\end{figure}

The age and the evolutionary status of the binary system components can be constrained using isochrones. We adopt isochrones computed with the PARSEC code (Bressan et al. 2012). The isochrone (its age) was selected to minimize the $\chi^2$ function including luminosities, effective temperatures and radii of the two components. Since we do not have a metallicity determination for the system, several sets of isochrones for different metallicity were considered. The best match was obtained for ${\rm [Fe/H]}=-0.45$, while we got a plausible fit for all ${\rm [Fe/H]}$ between -0.4 and -0.5. These values agree very well with metallicity of other eclipsing binary systems from LMC we have analyzed recently (Pietrzynski et al. 2013). The isochrone for ${\rm [Fe/H]}=-0.45$ is presented in the HR diagram in Fig.~\ref{fig:isochrone}. This Figure also shows the linear edges of the first overtone instability strip computed with the Warsaw convective pulsation codes (Smolec \& Moskalik 2008) (dashed lines, see Baranowski et al. 2009 for details). The selected isochrone corresponds to $\log age=8.266$, $\sim 185$ million years, with only a weak dependence on metallicity or mass loss rate (neglected in the presented isochrone). Both components are in a core-helium burning phase. The more massive Cepheid component is evolutionary more advanced and located at the end of the blue loop, well within the instability strip for first overtone Cepheids.

\section{Conclusions}
\label{sect:concl}
We have presented an analysis of the detached eclipsing binary Cepheid OGLE-LMC-CEP-2532, using the same methodology as for the OGLE-LMC-CEP-0227 system (P13). The faintness of the system and the lack of a secondary eclipse in the light curve imposed some simplifications in the analysis.
Despite these inconveniences we were able to derive the most important parameters of the Cepheid and the secondary component with good accuracy. The mass determination with an error of about $2.5\%$ is still one of the best measurements for Cepheid stars so far. It will serve well to solve the so-called Cepheid mass discrepancy problem and will allow to study the physics of classical Cepheids in general, as well as derive the relations connecting the mass of Cepheids with other parameters.

To test our model we have calculated the expected radius from the period-radius relations taken from Sachkov (2002) for fundamental and first-overtone mode Cepheids. We obtain radius values of $20.67 \pm 0.50 R_{\odot}$ and $26.8 \pm 3.9 R_{\odot}$, respectively. Our determination of the Cepheid radius ($28.95 \pm 1.4 R_{\odot}$) is consistent with the latter, which confirms that the star is a first-overtone pulsator as already indicated by the Fourier decomposition. It also shows that the degeneracy between $R_1$ and $R_2$ was resolved well during the analysis.

The radius change of the first overtone Cepheid analyzed here has a very small amplitude (about $1 R_{\odot}$) as compared to the fundamental mode Cepheid in the OGLE-LMC-CEP-0227 system (about $4 R_{\odot}$). Together with the faintness of the OGLE-LMC-CEP-2532 and the corresponding lower quality of the data, it was impossible to derive the $p$-factor for the Cepheid in this system.

Although good quality results were obtained this star is still an interesting object to follow. To better constrain the parameters and have a more accurate estimate of the $p$-factor more spectra have to be collected to better cover both the pulsational and the orbital radial velocity curve, but because of the long 800-day period this will take a few more years at least. Additional eclipse photometry over a wide wavelength
range is also highly desirable.  The secondary eclipse is probably too shallow to detect but with more measurements of better quality at the exact phase tighter constraints would be set on the inclination and the sum of the radii.

\acknowledgments
We gratefully acknowledge financial support for this work from the Polish National Science Center grant MAESTRO 2012/06/A/ST9/00269, from the BASAL Centro de Astrof{\'i}sica y Tecnolog{\'i}as Afines (CATA) PFB-06/2007 and from the Polish National Science Centre grant No. DEC-2011/03/B/ST9/02573. WG and DG also acknowledge support for this work from the Chilean Ministry of Economy, Development and Tourism's Millenium Science Initiative through grant IC120009 awarded to the Millennium Institute of Astrophysics (MAS); AG acknowledges support from FONDECYT grant 3130361. RS is supported by the Polish NSC grant UMO-2011/01/M/ST9/05914.

This paper utilizes public domain data obtained by the MACHO Project, jointly funded by the US Department of Energy through the University of California, Lawrence Livermore National Laboratory under contract No. W-7405-Eng-48, by the National Science Foundation through the Center for Particle Astrophysics of the University of California under cooperative agreement AST-8809616, and by the Mount Stromlo and Siding Spring Observatory, part of the Australian National University.

We would like to thank the support staff at the ESO Paranal and La Silla observatory and at the Las Campanas Observatory for their help in obtaining the observations and the rest of the OGLE team for their contribution in acquiring the data for the object. We thank ESO and the CNTAC for generous allocation of observing time for this project.

This research has made use of NASA's Astrophysics Data System Service.

{\it Facilities:} \facility{ESO:3.6m (HARPS)}, \facility{ESO:VLT (UVES)}, \facility{Magellan:Clay (MIKE)}, \facility{Warsaw telescope}.

\end{document}